\documentclass[twocolumn,prb,showpacs,floatfix,amsfonts,superscriptaddress]{revtex4} 
 
\usepackage[]{graphicx} 
\usepackage{amsbsy} 

\begin{document}
\title{Tuning a Schottky barrier in a photoexcited topological insulator with transient Dirac cone electron-hole asymmetry}
\author{M.~Hajlaoui}
\affiliation{Laboratoire de Physique des Solides, CNRS-UMR 8502, Universit\'{e} Paris-Sud, F-91405 Orsay, France}
\author{E.~Papalazarou}
\affiliation{Laboratoire de Physique des Solides, CNRS-UMR 8502, Universit\'{e} Paris-Sud, F-91405 Orsay, France}
\author{J.~Mauchain}
\affiliation{Laboratoire de Physique des Solides, CNRS-UMR 8502, Universit\'{e} Paris-Sud, 
F-91405 Orsay, France}
\author{L.~Perfetti}
\affiliation{Laboratoire des Solides Irradi\'{e}s, Ecole Polytechnique-CEA/DSM-CNRS UMR 7642,  91128 Palaiseau, France} 
\author{A.~Taleb-Ibrahimi}
\affiliation{Synchrotron SOLEIL, Saint Aubin BP 48, F-91192 Gif-sur-Yvette, France} 
\author{F.~Navarin}
\affiliation{Laboratoire de Physique des Solides, CNRS-UMR 8502, Universit\'{e} Paris-Sud, F-91405 Orsay, France}
\author{M.~Monteverde}
\affiliation{Laboratoire de Physique des Solides, CNRS-UMR 8502, Universit\'{e} Paris-Sud, F-91405 Orsay, France}
\author{P.~Auban-Senzier}
\affiliation{Laboratoire de Physique des Solides, CNRS-UMR 8502, Universit\'{e} Paris-Sud, F-91405 Orsay, France}
\author{C.R.~Pasquier}
\affiliation{Laboratoire de Physique des Solides, CNRS-UMR 8502, Universit\'{e} Paris-Sud, F-91405 Orsay, France}
\author{N.~Moisan}
\affiliation{Laboratoire d'Optique Appliqu\'{e}e, ENSTA ParisTech, CNRS, Ecole Polytechnique, 91761 Palaiseau, France}
\author{D.~Boschetto}
\affiliation{Laboratoire d'Optique Appliqu\'{e}e, ENSTA ParisTech, CNRS, Ecole Polytechnique, 91761 Palaiseau, France}
\author{M.~Neupane}
\affiliation{Department of Physics, Princeton University, Princeton, NJ 08544, USA}
\author{M.Z.~Hasan}
\affiliation{Department of Physics, Princeton University, Princeton, NJ 08544, USA}
\author{T.~Durakiewicz}
\affiliation{Los Alamos National Laboratory, Los Alamos, NM 87545, USA}
\author{Z.~Jiang}
\affiliation{School of Physics, Georgia Institute of Technology, Atlanta, Georgia 30332, USA}
\author{Y.~Xu}
\affiliation{Department of Physics, Purdue University, West Lafayette, IN 47907, USA}
\author{I.~Miotkowski}
\affiliation{Department of Physics, Purdue University, West Lafayette, IN 47907, USA}
\author{Y.P.~Chen}
\affiliation{Department of Physics, Purdue University, West Lafayette, IN 47907, USA}
\author{S.~Jia}
\affiliation{Department of Chemistry, Princeton University, Princeton, NJ 08544, USA}
\author{H.W.~Ji}
\affiliation{Department of Chemistry, Princeton University, Princeton, NJ 08544, USA}
\author{R.J.~Cava}
\affiliation{Department of Chemistry, Princeton University, Princeton, NJ 08544, USA}
\author{M.~Marsi}
\affiliation{Laboratoire de Physique des Solides, CNRS-UMR 8502, Universit\'{e} Paris-Sud, F-91405 Orsay, France}
\date{\today}

%

\maketitle

\textbf{
The advent of Dirac materials has made it possible to realize two dimensional gases of relativistic fermions with unprecedented transport properties in condensed matter. Their photoconductive control with ultrafast light pulses is opening new perspectives for the transmission of current and information. 
\\
Here we show that the interplay of surface and bulk transient carrier dynamics in a photoexcited topological insulator can control an essential parameter for photoconductivity - the balance between excess electrons and holes in the Dirac cone. This can result in a strongly out of equilibrium gas of hot relativistic fermions, characterized by a surprisingly long lifetime of more than 50 ps, and a simultaneous transient shift of chemical potential by as much as 100 meV. 
The unique properties of this transient Dirac cone make it possible to tune with ultrafast light pulses a relativistic nanoscale Schottky barrier, in a way that is impossible with conventional optoelectronic materials.  
}

Low energy electron bands in graphene~\cite{Geim2007} and surface states of 3D Topological Insulators (TI's)~\cite{Hasan2010,Qi2011} can be described by 2D Dirac equations, and are the most promising candidates for a new generation of devices based on relativistic fermions~\cite{Pesin2012}. These Two-Dimensional Dirac Systems (2DDS's) represent a novel state of quantum matter characterized by revolutionary transport properties: some of these features, like the unprecedented mobility of graphene~\cite{Geim2007} or the immunity to defects and impurities of TI's~\cite{Hasan2010,Qi2011}, make Dirac fermions particularly interesting for ultrafast photonics and optoelectronics~\cite{Xia2009,Sun2012}. Their applicability to real devices requires the ability to control the lifetime and population of hot photocarriers, and to manipulate the balance between excess electrons and holes in the transient regime.  

Several studies have already been performed to explore the response of 2DDS's to ultrafast light pulses, but none of them has focused on the effects of an asymmetric electron-hole population in the non-equilibrium Dirac cone. In the case of graphene,  electron-hole symmetry  has been consistently found in the photoexcited state, and it naturally stems from the direct intra-cone nature of the optical excitation~\cite{Winnerl2011,Li2012,Gilbertson2012}. In TI's direct optical excitation has been shown for a high-lying unoccupied surface state~\cite{Sobota2013}, but for the topological surface state an excess carrier population can be instead built up in the Dirac cone via interband scattering processes involving also bulk valence and conduction states~\cite{Sobota2012, Hajlaoui2012, Wang2012, Crepaldi2012,Sobota2013,Crepaldi2013}. This necessarily leads to the additional problem of a strong presence of hot excess carriers in the bulk bands during the whole transient regime following the optical excitation~\cite{Sobota2012, Hajlaoui2012, Wang2012, Crepaldi2012}. This is particularly disturbing in the perspective of using TI's for optoelectronic applications, because one does not preserve in the photoexcited regime the essential feature of a topological insulator - the coexistence of a gapless surface state and of a bulk insulator.  Here we show that these limitations can be overcome by considering an insofar neglected element: the action of the band bending on the interplay between surface and bulk carrier dynamics~\cite{Bokor1989, Halas1989, Marsi1997, Toben2005}. If correctly mastered, the band lineup can be advantageously used to control and modify the charge balance and transient population of hot Dirac fermions in a 3D TI, while keeping at the same time the subsurface space charge layer as void as possible of excess carriers. In this way, the 2DDS constitutes a strongly out of equilibrium conducting state that cannot be obtained with a conventional metal, with interesting implications for optoelectronic devices based on topological insulators. 

\section*{Results}

\begin{center}
\begin{figure*}[htb]
\includegraphics[width=1\linewidth,clip=true]{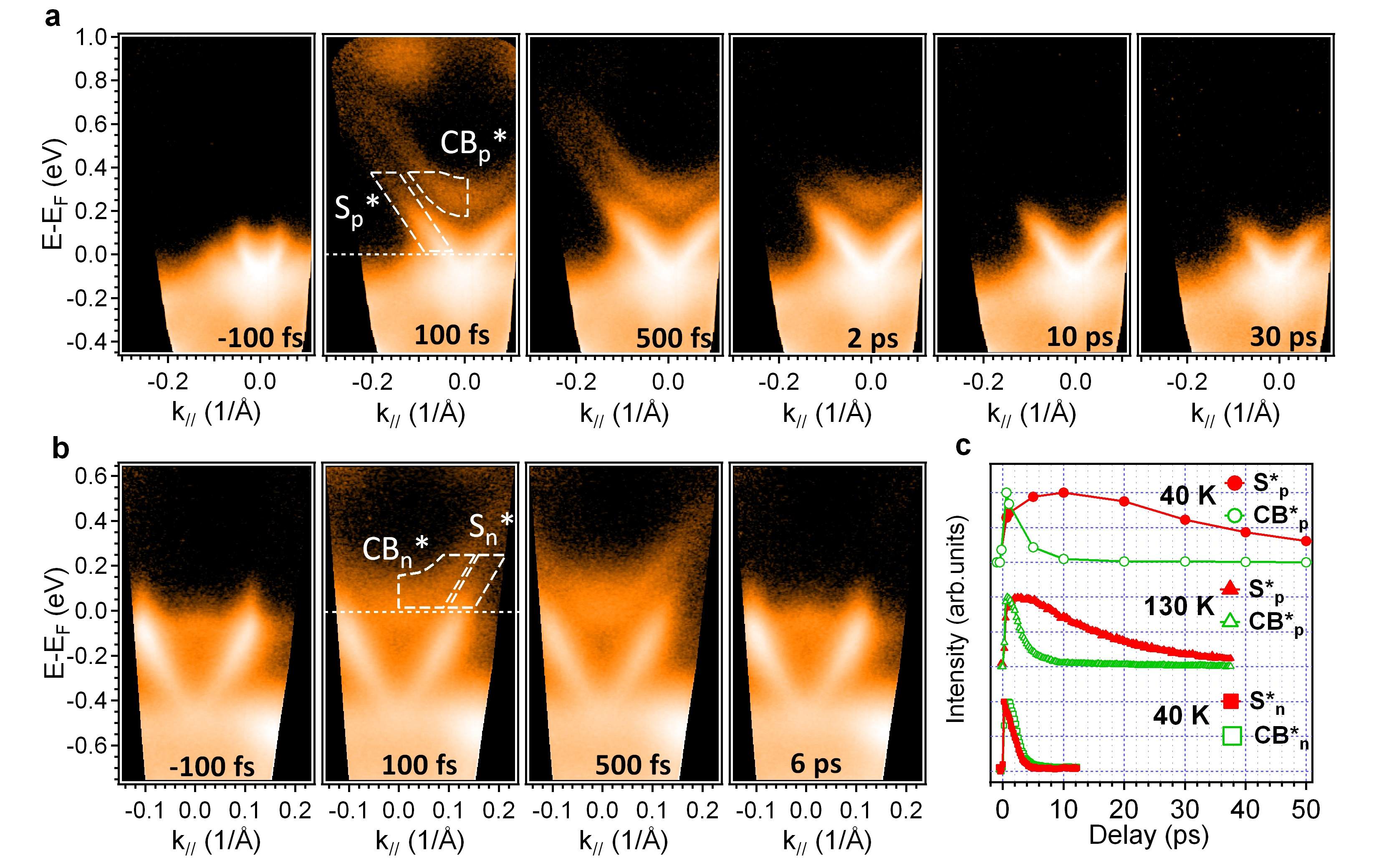}
\caption{
{\bf Dirac cone ultrafast dynamics detected with time-resolved ARPES}. Selected time delays are presented for (a) p-type Bi$_{2.2}$Te$_3$ and (b) n-type Bi$_2$Te$_2$Se. The signal intensity is presented in a logarithmic color scale. The integration regions called S$_p$*, S$_n$*, and CB$_p$*, CB$_n$* are indicated as dashed contour windows. (c) time evolution of S$_p$*, S$_n$*, CB$_p$*, CB$_n$*. 
}
\label{Fig1}
\end{figure*}
\end{center}

\begin{center}
\begin{figure*}[htb]
 \includegraphics[width=1\linewidth,clip=true]{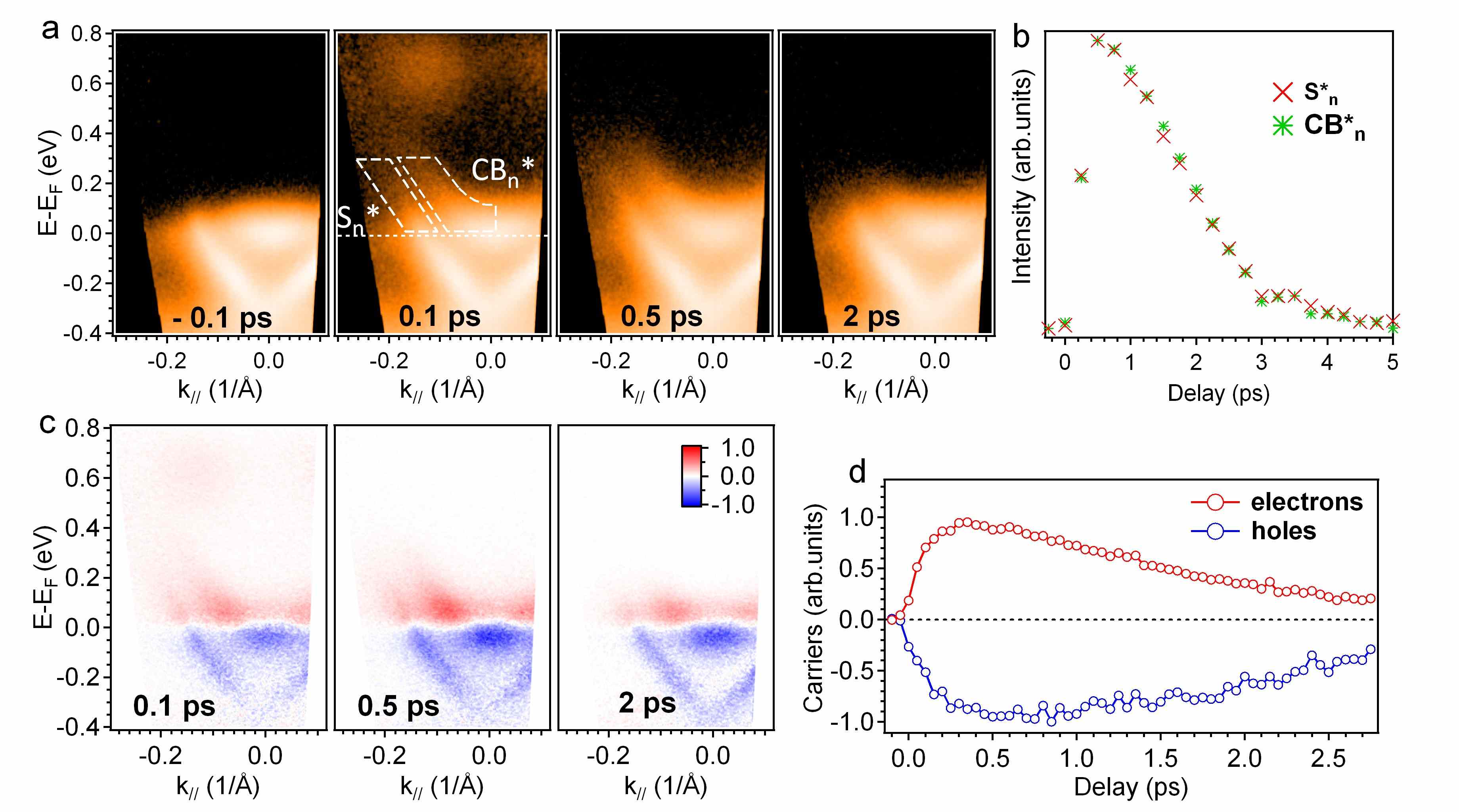}
\caption{
{\bf Ultrafast dynamics of a flat-band, n-type Bi$_2$Te$_3$ sample}. 
(a) Time-resolved ARPES images. (b) Time evolution of the electron 
populations in the Dirac cone and in the conduction band: as one can see, the relaxation time for excess electrons
in the Dirac cone is much faster than for the p-type sample (see also Supplementary Figure S1), because in this 
case they are strongly coupled with the conduction band and they have access to the faster bulk 
carrier recombination channels. 
(c) $k-E$ distribution of excess electrons and holes; 
(d) time evolution of the excess electron and hole populations.  
As expected for a flat band system, the electron and hole
populations are very well balanced for all time delays: this becomes even more evident by 
comparing Fig. 2(c) and (d)  with Fig. 3, where 
asymmetric populations are detected for excess holes and electrons, respectively. 
}
\label{Fig2}
\end{figure*}
\end{center}

{\bf Lifetime of excited Dirac fermions.} We were able to unveil these effects through a series of time-resolved ARPES measurements performed on several bismuth chalcogenide TI's~\cite{Zhang2009, Yazyev2010, Hsieh2008, Chen2009}. As representative examples, we present in Fig.\ref{Fig1} images obtained for selected time delays $\Delta$t on p-type Bi$_{2.2}$Te$_3$, with a downward band bending of about 60 meV at the surface, and on Bi$_2$Te$_2$Se, which at 40 K is n-type~\cite{Jia2012,Xiong2012,Ji2012,Niesner2012,Mi2013} and presents an upward band bending of about 30 meV. For sake of comparison, in Fig.\ref{Fig2} we present 
results from a strongly n-doped Bi$_2$Te$_3$ sample presenting flat bands at the surface. 
We also provide the complete time-resolved ARPES sequences for all the specimens as movies: for p-type Bi$_{2.2}$Te$_3$ (Supplementary Movie 1); for n-type Bi$_2$Te$_2$Se (Supplementary Movie 2); and for strongly n-doped Bi$_2$Te$_3$ (Supplementary Movie 3). In Fig.\ref{Fig1}c we present the time evolution of the population in the $k-E$ windows CB$_p$*, CB$_n$* (the bulk conduction band) and S$_p$*, S$_n$* (the excited Dirac cones) as defined in Fig.\ref{Fig1}a and Fig.\ref{Fig1}b for  p-type Bi$_{2.2}$Te$_3$ and n-type Bi$_2$Te$_2$Se, respectively. As discussed in previous works~\cite{Sobota2012, Crepaldi2012}, decay rates depend on the specific momentum-energy region: after an analysis of this dependence for our samples, detailed in the Supplemenary Figure S2, we found that this choice for the $k-E$ windows allows us to present in Fig.\ref{Fig1}c the essential information we need for this study. 

During the first few picoseconds an excess carrier population accumulates in the Dirac cone, as a result of different interband scattering processes that also involve bulk states, as thoroughly described in previous studies~\cite{Sobota2012, Hajlaoui2012}. For longer delays, n-type Bi$_2$Te$_2$Se is almost completely relaxed after 8 ps, while in p-type Bi$_{2.2}$Te$_3$ a significant population of excess hot carriers survives in the surface Dirac cone with a strikingly long lifetime of $\tau_S$ $\approx$ 50 ps at T = 40 K. At 130 K, this lifetime is reduced (30 ps), but it is still considerably larger than observed in previous studies on TI's~\cite{Sobota2012, Hajlaoui2012, Wang2012, Crepaldi2012}. 

By focusing our attention on the p-type specimen, it is immediately clear that after the first few ps one has filled the available empty states in the Dirac cone within the gap. In Supplementary Figure S2 we discuss this in more detail, also by comparing the results with the n-type Bi$_{2}$Te$_3$ sample. It has already been observed~\cite{Sobota2012} that excited surface states have a longer relaxation time when they are within the gap, because they do not have access to the faster bulk recombination channels. In our case, this is very evident from the analysis of Supplementary Figure S1 (for p-type Bi$_{2.2}$Te$_3$) and of Fig.\ref{Fig2}b (for n-type Bi$_{2}$Te$_3$). 
Furthermore, the sequence in Fig.\ref{Fig1}a reveals a qualitatively different process with respect to previous works on p-type Bi$_{2}$Se$_3$, where a persisting non-equilibrium population of bulk photoinduced carriers CB$_p$* was found in the conduction band, acting as a charge reservoir for the Dirac cone states in the gap~\cite{Sobota2012, Crepaldi2012}. In our case, for $\Delta t \gtrsim$ 10 ps, the p-type Bi$_{2.2}$Te$_3$ sample still presents a strong excess electron population in the surface state, while no carriers can be detected at the bottom of the conduction band. 

\begin{center}
\begin{figure*}[htb]

 \includegraphics[width=1\linewidth,clip=true]{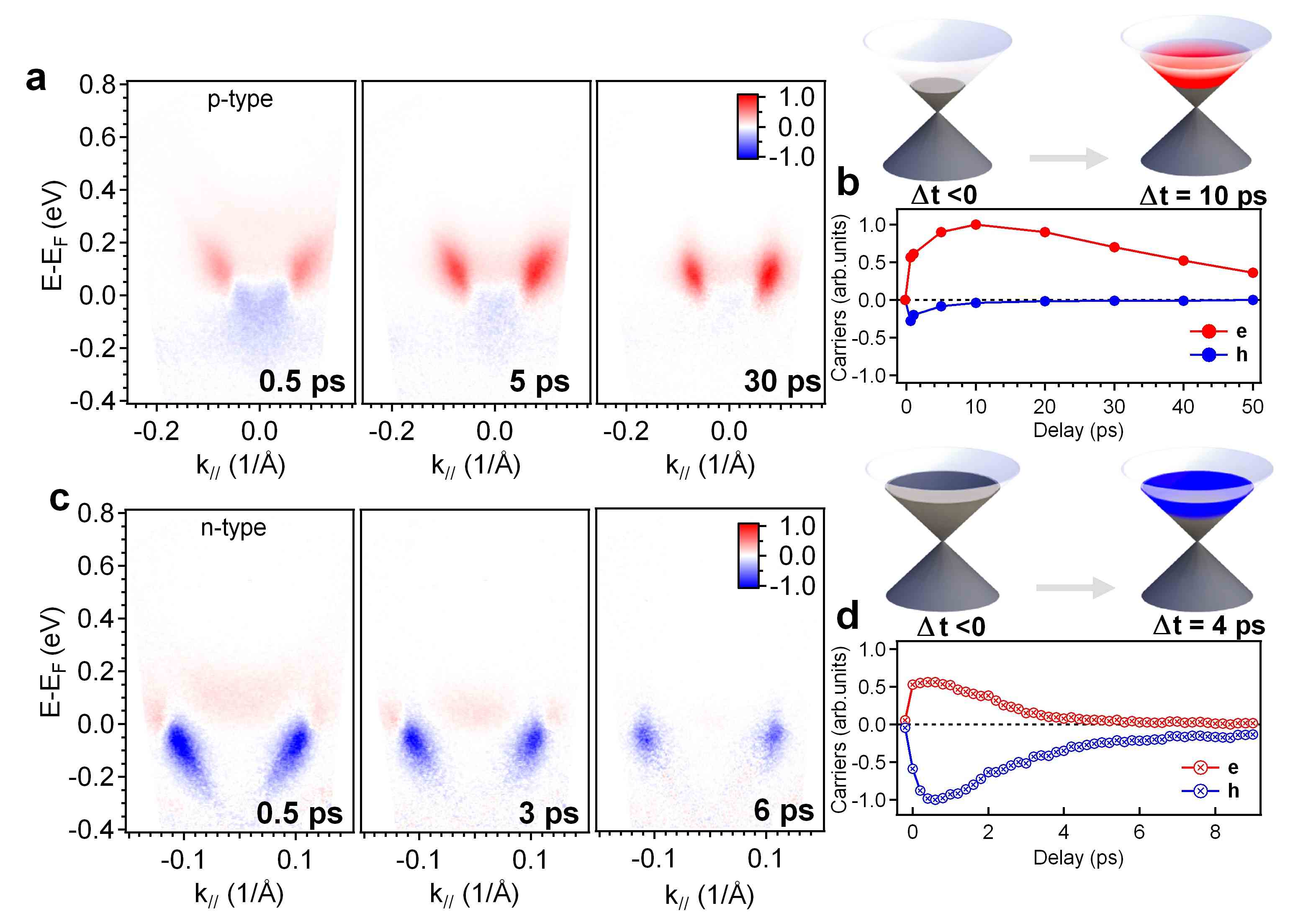}
\caption{
{\bf Charge asymmetry of transient excess carriers}. Excess electrons (red) and holes (blue) are detected from the difference of time-resolved ARPES images taken after and before photoexcitation. We present them in (a) for p-type Bi$_{2.2}$Te$_3$ and in (c) for n-type Bi$_2$Te$_2$Se. In (b) and (d) we present the time evolution of the detected excess carriers (electrons, e, and holes, h): a transient accumulation of holes is seen for the n-type specimen, and a much more pronounced asymmetry related to a strong and long lived accumulation of hot electrons in the Dirac cone for the p-type sample.
}
\label{Fig3}
\end{figure*}
\end{center}

{\bf Surface band bending and transient charge asymmetry.} The reason for this is that a preexisting surface band bending can strongly affect the bulk-surface interplay, and consequently the Dirac cone relaxation, by spatially separating excess electrons and holes. In our case, the inequivalent charge carrier dynamics can be extracted from the differences of the ARPES images with respect to the non-excited systems (Fig.\ref{Fig3}a and Fig.\ref{Fig3}c). In Fig.\ref{Fig3}b and Fig.\ref{Fig3}d we present the time evolution of the detected excess electron and hole populations. For the n-type sample, with bands slightly bent upwards, a clear excess of holes can be found in the Dirac cone for a few ps. The charge asymmetry is much more pronounced for the p-type specimen. After 500 fs the hole population starts decreasing in the surface layers (with 6.3 eV photons our ARPES probing depth is about 2-3 nm~\cite{Rodolakis2009}) while excess electrons maintain a strong presence as they are actually trapped in the surface states (see also Fig.\ref{Fig3}a). This is due to the surface band bending, which induces a flow of the excess majority carriers (holes) towards the bulk, while the excess minority carriers (electrons) remain confined at the surface. This charge disequilibrium at the surface explains the unusually long relaxation time for  Bi$_{2.2}$Te$_3$ (excess electrons find no holes to recombine with), as well as its temperature dependence (the bulk mobility of the carriers is lower at 40 K than at 130 K due to impurity scattering). The corresponding transient surface photovoltage~\cite{Kronik1999, Marsi1998} was measured for the various time delays (Supplementary Figure S3) and found to be negligibly small ($\approx$ 10 meV) for the laser fluences used in this study. 

{\bf Electron-hole asymmetry as a carrier recombination bottleneck.} The lack of holes in the surface and subsurface region acts as the effective bottleneck for the recombination of excess Dirac electrons, which present an unusually long lifetime for the basic reason that there are not enough positive charge carriers to recombine with. If the bands are instead flat, the distribution of excess bulk electrons and holes is essentially always balanced (in time t and in space z), and the recombination of excess Dirac carriers will be mainly limited by intrinsic factors, like the electron-phonon scattering rate. 
This is indeed the case for the flat-band n-type Bi$_{2}$Te$_3$ sample presented in Fig. 2, as well as 
for a prototype system where no surface-bulk interplay is present, namely non-equilibrium Dirac cones in graphene~\cite{Winnerl2011, Li2012}. In the case of 3D TI's, various time-resolved ARPES studies have examined how the weak electron-phonon coupling determines the slow relaxation of the surface states: in particular the sibling compound Bi$_2$Se$_3$ has already been extensively explored~\cite{Sobota2012, Wang2012, Crepaldi2012, Crepaldi2013} for different temperatures and for pump fluences  ranging from 0.025 mJ cm$^{-2}$ to 0.25 mJ cm$^{-2}$, with relaxation rates for the Dirac cone excess population between 5 and 10 ps.  Here we are confronted with a markedly different time scale for the Dirac cone relaxation (of the order of 50 ps or more, and strongly dependent on the sample temperature) which is due to a different underlying physical process, namely the transient macroscopic charge carrier separation. This is unambiguously confirmed by the fact that the relaxation time was found to increase even more when the bands were more bent. We were able to perform this test by letting a cleaved surface exposed to the residual gases of our ultrahigh vacuum system: consistently with previous studies ~\cite{Hsieh2009,Bianchi2010} after 24 hours we observed an increase of surface band bending, and we found that the Dirac cone relaxation rate increased as well (see Fig. 4). This is possible thanks to the well known robustness of the surface of topological insulators and indicates possible practical pathways to actively control this effect. 

{\bf Transient topological insulator state.} In Fig.\ref{Fig5} we illustrate the consequences of the spatial separation of photoexcited negative and positive carriers for the case where they are more pronounced and long lasting, i.e. for p-type Bi$_{2.2}$Te$_3$ (specular arguments may be applied to the case of n-type Bi$_2$Te$_2$Se). 
The photoexcited excess electrons, confined to the sub-surface region, fill the surface Dirac cone over few ps, and, after 10 ps, there are no visible electrons in the conduction band. There are also no excess holes in the subsurface valence band, because they have drifted deeper into the bulk ($z \gtrsim L_D$, where $L_D$ is the Debye length of the system); furthermore, for $0 \leq z \lesssim L_D$ the Fermi level is really within the band gap. Consequently, within the space charge layer (i.e. the subsurface region corresponding to $0 \leq z \lesssim L_D$) a situation is created where there is a strong excess population of electrons concentrated in the conducting surface states while no excess carriers are present in the bulk bands: therefore, the space charge layer preserves also in the transient regime this main distinctive feature of TI's. If one gates the system on a time window starting 10 ps after the pump pulse, this portion of the specimen can consequently be regarded as a genuine transient topological insulator, whose extension in space ($L_D \approx$ 20 nm) and in time ($\Delta t \approx$ 50 ps) can be interesting for ultrafast optoelectronic applications. 

{\bf Phototuning of a relativistic Schottky barrier.} In order to appreciate the character of this transient state, we can consider it as a nanometer scale metal-semiconductor junction between the metallic 2DDS and the subsurface space charge layer. In its ground state the chemical potentials $\mu$ of the 2DDS and of the substrate are in equilibrium, and the energy difference between the top of the valence band and $\mu$ can be regarded as the height $\phi$ of the Schottky barrier (Fig.\ref{Fig5}). The transient filling of the Dirac cone after photoexcitation can be regarded as a shift of a chemical potential defined only for the surface 2DDS, $\mu$*. In the transient topological insulator state ($\Delta t \gtrsim$ 10 ps), the substrate bands are virtually the same as in the ground state, but the barrier height of the Schottky barrier has changed by about 0.1 eV, i.e. more than half the gap of Bi$_{2.2}$Te$_3$. Even more remarkably, this large change is only due to the shift of $\mu$* in the 2DDS, which in this transient regime is not in equilibrium with the substrate. It should be emphasized that for junctions between a semiconductor and a normal metal the Schottky barrier height can be changed in many ways, but these methods always involve changing the lineup of the semiconductor bands, while the metal side is unaffected~\cite{Sze}. In the current case, it is instead the metallic side that is strongly perturbed: this is a consequence of the unique properties of the 2DDS's at the surface of TI's, namely the fact that they simultaneously present a metallic nature and an exceptionally long relaxation time for their photoexcited state.   

{\bf Analysis of the non-equilibrium Dirac cone.} A detailed analysis of the Energy Distribution Curves (EDC's) gives a more complete description of the evolution of this strongly out of equilibrium Dirac cone. 
In Fig.\ref{Fig6}a we consider momentum-selected EDC's taken along the surface bands of p-type Bi$_{2.2}$Te$_3$ at T=130 K, and we present in the contour plot the difference of the photoexcited Dirac cone with respect to its equilibrium state. In Fig.\ref{Fig6}c we present the EDC's in logarithmic scale for two limiting cases, $\Delta$t $<$ 0 (non excited state) and 0.5 ps (strongly out of equilibrium state). The momentum-selected EDC for $\Delta$t $<$ 0 can be well fitted with a Fermi-Dirac distribution, with T$_e$=130 K, convoluted with our experimental resolution (80 meV). The spectrum for $\Delta$t = 0.5 ps presents a non zero signal level at high kinetic energies, a background due to incoming scattered electrons. To model the EDC, we describe this background with a constant energy distribution $D_2$* for E $\geq$ $E_F$, plus a Fermi-Dirac distribution with temperature T$_e$ and chemical potential $\mu$* to describe the rest of the excess population $D_1$*. 
This model makes the assumption that an electronic temperature T$_e$ can be defined for a subset of the overall electron population, neglecting $D_2$*, and it provides a good fit for the spectra at all delays (as it can be seen for selected time delays in Fig.\ref{Fig6}c). The parameters extracted from the fit are plotted in Fig.\ref{Fig6}b, and indicate that a real electronic thermalization develops only when $D_2$*=0 and when $\mu$* reaches its maximum value (for $\Delta$t = $\tau_t$, with $\tau_t$ = 4 ps at T = 130 K, Fig.\ref{Fig6}d). Strictly speaking, when $D_2$*$\neq$0 parameters like $\mu$* and T$_e$ are physically ill-defined, and we report them in Fig.\ref{Fig6}b with different symbols for the sake of clarity. This distinction is particularly clear here because the band filling ($\Delta$t $\leq$ $\tau_t$) and its quasi-adiabatic relaxation ($\Delta$t $\geq$ $\tau_t$) take place on well distinct time scales~\cite{Lisowski2004}. Hence, for the transient topological insulator state ($\Delta$t $\geq$ 10 ps) a quasi-equilibrium chemical potential for the Dirac cone can be rigorously defined; this corresponds to the $\mu$* introduced in Fig.\ref{Fig5}.

\section*{Discussion}

From the fundamental point of view, this state represents a model case of universal relevance for a non-equilibrium fermion distribution, and shows that electron-hole asymmetry in the subsurface carrier population can be a more stringent factor in the relaxation of a hot 2DDS than intrinsic processes like electron-phonon scattering. 
It should be emphasized that only the combination of the unique intrinsic properties of the 2DDS and of its interplay with the semiconductor substrate make it possible to drive a metallic state in such a markedly out of equilibrium state. The key for achieving this condition is that the intrinsically slow recombination time for the surface states is long enough to give subsurface excess majority carriers the time to drift into the bulk. For exemple, for the case of the p-type specimen discussed in this paper, the time necessary to create a strong electron-hole asymmetry in the subsurface region and in the surface bands (few ps, see Fig.\ref{Fig3}a and Fig.\ref{Fig5}b ) is faster that the intrinsic Dirac cone relaxation, primarily determined by electron-phonon scattering. At this point (approximately at t=$\tau_t$) the still very high number of excess electrons in the Dirac cone are left with an insufficiently small number of holes to recombine with: since the total electrical charge must be conserved during the recombination process, the asymmetry between electrons and holes becomes the most stringent bottleneck for their recombination, which is consequently further slowed down to a considerably longer time scale (of the order of 50 ps). The necessity of this simultaneous combination of different factors explains why such a long-lived, strong shift of the chemical potential has never been observed for a metallic system (neither for a metallic film at a junction with a semiconductor, nor for a 3D metal). 

From the point of view of possible technological applications, it is well known that a precise control of the chemical potential position is a key issue and one of the main challenges towards the realization of materials with genuine TI properties~\cite{Hasan2010,Qi2011}, i.e. systems where the chemical potential is in the middle of the bulk bandgap, the Dirac cone is partially occupied and there are no excess carriers in the bulk bands. The transient state described in this paper does present all these properties, and it demonstrates that manipulating TI's with ultrafast light pulses can provide alternative and powerful pathways for the control of their unique transport properties.

In particular, we have observed that for an electron (or hole) travelling along the z direction, the 2DDS at the surface of a 3D TI represents a Schottky barrier. As we have discussed, an advantageous time window exists in the transient regime such that the space charge layer region $0 \leq z \lesssim L_D$ can be found in virtually the same state as before photoexcitation, while the 2DDS presents a photo-shifted chemical potential: this system represents a potentially interesting photo-excited TI, where the Schottky barrier height can be transiently tuned with ultrafast light pulses. This can be used for instance in a new type of transiently tuneable Schottky diodes based on Dirac fermions, by selectively acting with nanoscale electrical contacts only on the space charge layer region. The possibility we described of maintaining a genuine TI also in the photoexcited state is of relevance for any photoconductive application of this kind of 2DDS, especially if one considers that the thickness of the space charge layer region (around 20 nm) is comparable to the penetration length of optical photons in the material. In general, our findings are of interest for any kind of applications where one wants to use light to transfer electrons from the semiconducting substrate to the surface 2DDS, or conversely if one wants to prevent this transfer. 

In conclusion, our results indicate a general approach to effectively photoinduce a strongly out of equilibrium state by acting on the charge balance in the Dirac cone of TI's. This photoexcited 2DDS simultaneously presents a metallic nature and an extremely long recombination time; it may for instance be used in a nanometer scale Schottky photodiode, because the 2DDS is simultaneously atomically thin and strongly, topologically decoupled from the substrate. All these properties cannot be found in a conventional metal, and indicate many interesting perspectives for the ultrafast photoconductive control of 2DDS's. In the case of graphene, a strong photoinduced electron-hole asymmetry could be produced with the use of high energy photons, involving indirect scattering with higher energy bands: this could be very interesting in combination with the realisation of graphene based metal-semiconductor contacts~\cite{Hicks2013}. For TI's, additional opportunities are related to the possibility of actively exploiting the helical spin texture of the Dirac fermions in  optospintronics. The photon energy and the electric field related to band bending, along with the relative position of the Fermi level and of the Dirac point, are the main factors determining the properties of the excited Dirac cone. Since all these factors can be tailored at will using adjustable parameters, they appear to be essential tools for future devices based on the photoconductive control of Dirac fermions.

\section*{Methods}

{\bf Transport measurements.} The high quality single crystals used for these experiments were thoroughly characterized prior to the time-resolved ARPES experiments. The growth of the Bi$_2$Te$_2$Se crystals employed in this study had been described elsewhere~\cite{Jia2012}, as has their basic electronic characterization~\cite{Jia2012,Xiong2012}. 
The Bi$_{2.2}$Te$_3$ and Bi$_{2}$Te$_3$ crystals were grown by the Bridgman method and their 
transport properties have been determined by measuring their longitudinal and transverse Hall resistivities. Six gold contacts were evaporated on the samples in a standard geometry. From these measurements, the resistivity and conductance tensors can be fully extracted after symmetrization (antisymmetrization) of the longitudinal (transverse) resistivities with respect to the magnetic field, which was applied perpendicular to the two layers. In addition, thermopower experiments were also performed on both materials as a function of temperature. The analysis of the Hall coefficients (Supplementary Figure S4) and Seebeck coefficients (Supplementary Figure S5) allowed us to conclude that the Bi$_{2}$Te$_3$ specimen was strongly n-doped, with a bulk carrier density $n \approx 10^{21}\:cm^{-3}$. The Bi$_{2.2}$Te$_3$ was instead p-doped, with a bulk hole carrier density $n_h\approx 9 \times 10^{18} cm^{-3}$. 

{\bf Time-resolved ARPES measurements.} The band llneup was precisely determined by comparing the bulk carrier concentration extracted from transport measurements with the band position observed with surface sensitive ARPES on clean, freshly cleaved specimens. 
The time-resolved ARPES measurements were performed on the FemtoARPES setup, in experimental conditions similar to our previous work~\cite{Hajlaoui2012}. The laser source was operated at 0.25 MHz repetition rate, pumping the specimens with 1.57 eV pulses (35 fs, 0.1 mJ cm$^{-2}$) and probing them with the 4th harmonic at 6.3 eV~\cite{Faure2012}, with an overall time resolution of 65 fs and an energy resolution of 80 meV. The data presented here were taken along $\Gamma$K, but we found no substantial differences for measurements performed along other directions in k-space on several cleaved surfaces, obtaining consistently similar results for the dynamics of the excited states.

\section*{Acknowledgements}

We thank M.O. Goerbig, J.-N. Fuchs and G. Montambaux for very interesting discussions. 
Materials synthesis at Purdue is supported by the DARPA MESO program (Grant N66001-11-1-4107). Z.J. acknowledges support from the DOE (DE-FG02-07ER46451). T.D. was funded by LDRD and BES programs at LANL, under the auspices of the DOE for Los Alamos National Security LLC and by Office of Basic Energy Sciences, Division of Material Sciences. M.N. and M.Z.H. are supported by NSF-DMR-1006492. Crystal growth and electronic characterization in Princeton were supported by US DARPA grant N6601-11-1-4110. The FemtoARPES activities were funded by the RTRA Triangle de la Physique, the Ecole Polytechnique, the EU/FP7 under the contract Go Fast (Grant No. 280555), the ANR (Grant ANR-08-CEXCEC8-011-01) and the Labex PALM.

\section*{Author contributions} 

M.H., E.P., J.M. and L.P. performed the time-resolved ARPES experiments and data analysis, 
with contributions from A.T., M.Z.H., T.D. and R.J.C. N.M, D.B., M.N., performed complementary 
measurements, while F.N., M. Mo., P.A.S. and C.R.P. collected and analyzed the transport data. 
Y.X., I.M., Y.P.C., S.J., H.W.J. and R.J.C. grew the samples. M.Ma. provided the overall project coordination. 
All authors contributed to the discussion and the writing of the paper. 

\section*{Competing financial interests} 

The authors declare no competing financial interests.

\section*{Contact information} 

Corresponding author M. Marsi ; email address:    marino.marsi@u-psud.fr

\begin{figure*} [t]
\includegraphics[width=0.7\linewidth,clip=true]{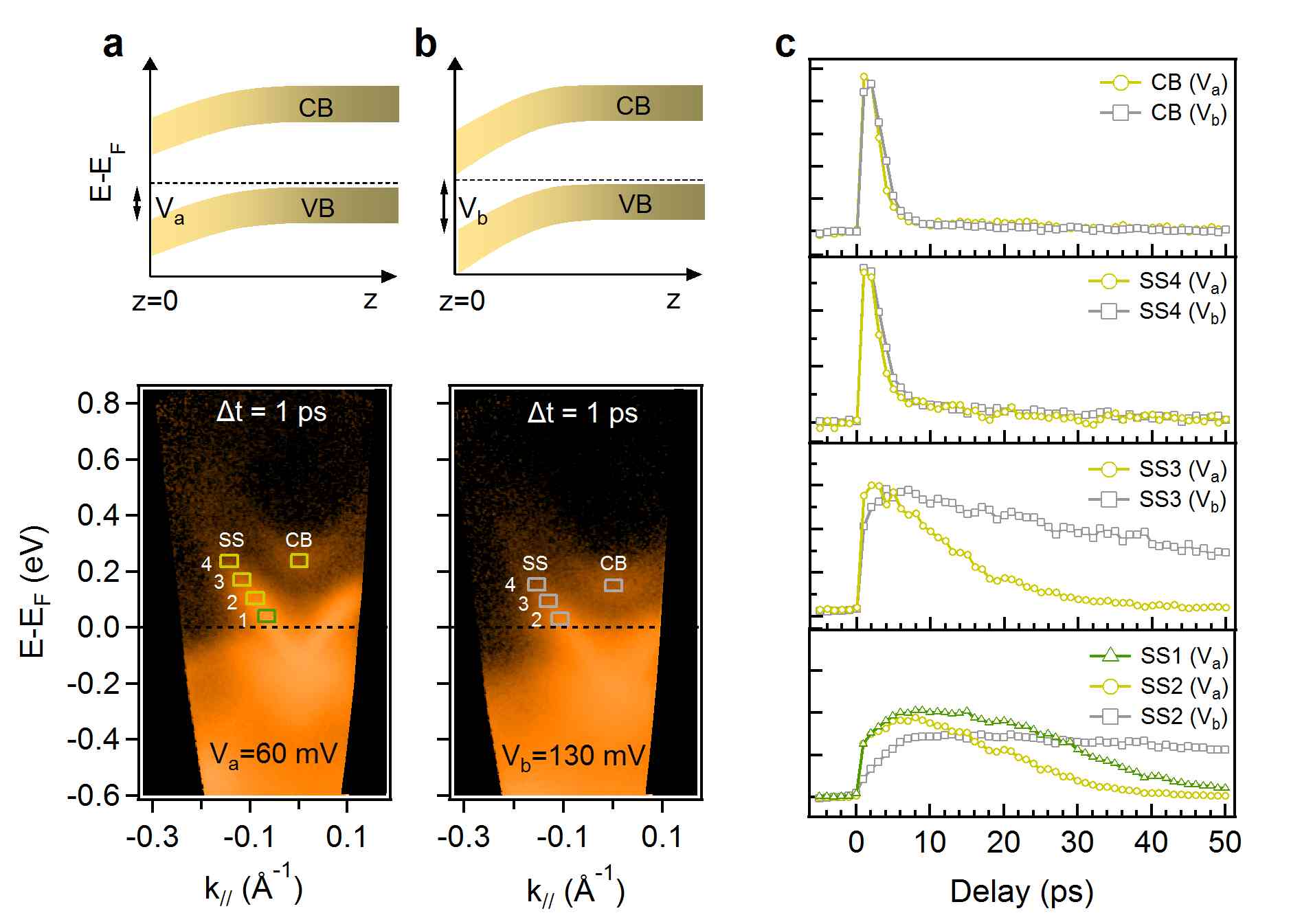}
\caption{
{\bf Effect of band bending on the recombination rate for p-type Bi$_{2.2}$Te$_3$} 
Decay rates for photoexcited p-Bi$_{2.2}$Te$_3$ at T=40 K in the case of (a) a V$_a$=60 mV 
band bending 
and (b) of a V$_b$=130 mV band bending (b). The pump-probe ARPES images  
correspond to a
time delay of 1 ps.  In (c) the decay rates for equivalent energy windows are compared. 
The energy windows for the surface state SS have been chosen so that SS4 has from the same 
binding energy as the bottom of the conduction band CB. As it can be clearly seen, equivalent windows for the 
two surfaces show a much longer relaxation time for surface (b), where the band bending is more 
pronounced. In particular, SS2 for the case (b) has a lifetime $\geq$ 100 ps  (there is of course no equivalent in 
(b) of window SS1 because, due to the increased band bending, in (b) SS1 becomes part of the filled electronic 
states). 
}
\label{Fig4}
\end{figure*}

\begin{figure*}[t]
  
\includegraphics[width=1\linewidth,clip=true]{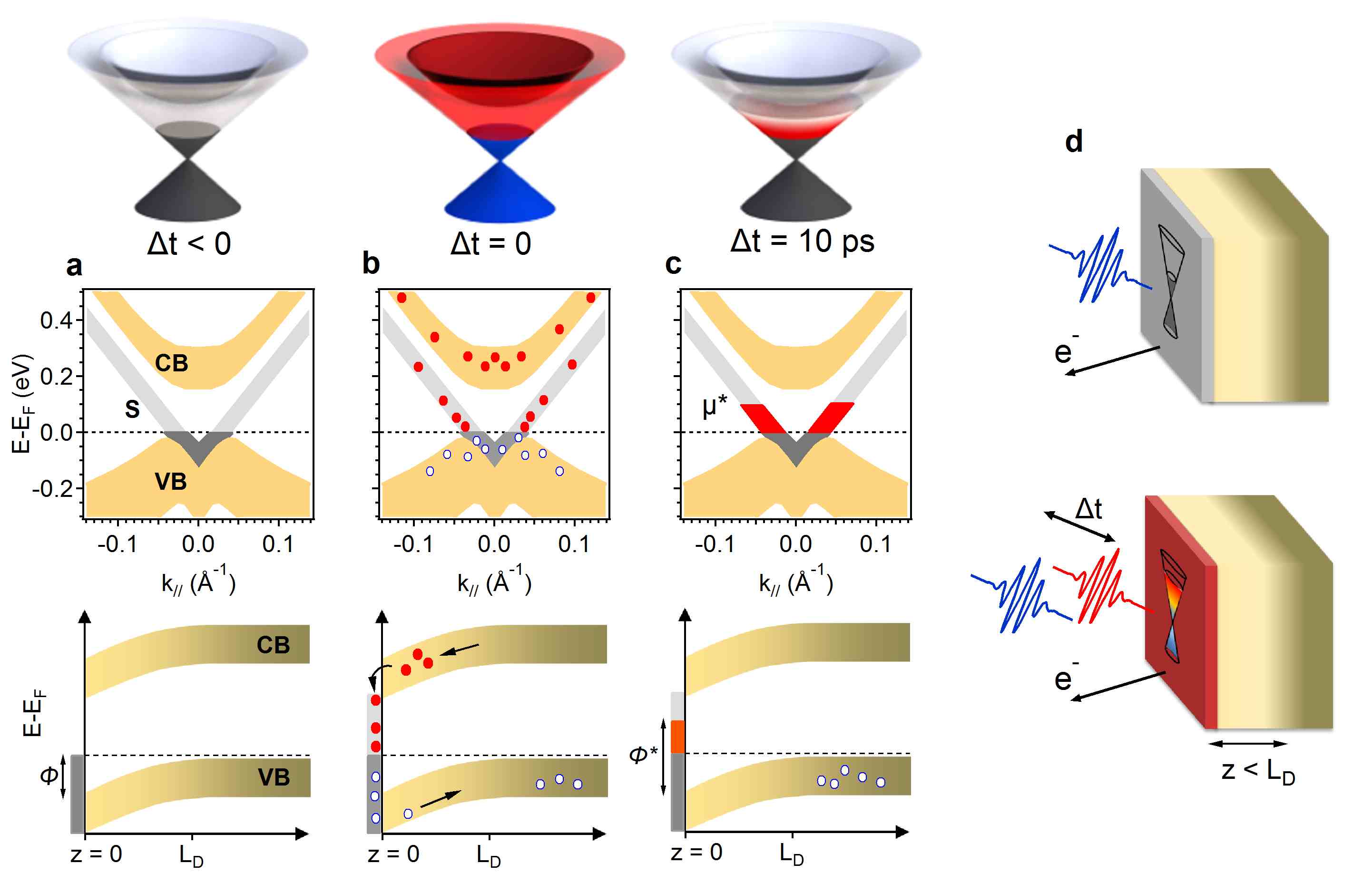}
\caption {
{\bf Transient topological insulator state and non equilibrium surface 2D metallic layer in a nanoscale Schottky barrier}. We compare Bi$_{2.2}$Te$_3$ (a) in its ground state, (b) right after photoexcitation and (c) in its photo-excited state for a delay $\Delta t$ = 10 ps after optical pumping. For each case, the top panel represents a pictorial view of the Dirac cone with excess electrons (red) and holes (blue); the mid panel shows the schematic structure of the bands in reciprocal space and their filling at the surface (z=0); and the bottom panel the corresponding band lineup in real space along the normal to the surface z. In (d) we present a schematic view of the experimental configuration: when the sample is in its ground state (top), ARPES probes the rest value $\phi$ of the barrier between the chemical potential $\mu$ and the top of the valence band; when the sample is photoexcited with a pump pulse, time resolved ARPES probes the evolution of the transient topological insulator state vs time delay $\Delta$t (bottom): the difference of the Schottky barrier height $\phi$*-$\phi$ is only due to the strongly out of equilibrium 2DDS.
}
\label{Fig5}
\end{figure*}

\begin{figure*}
 
 \includegraphics[width=1\linewidth,clip=true]{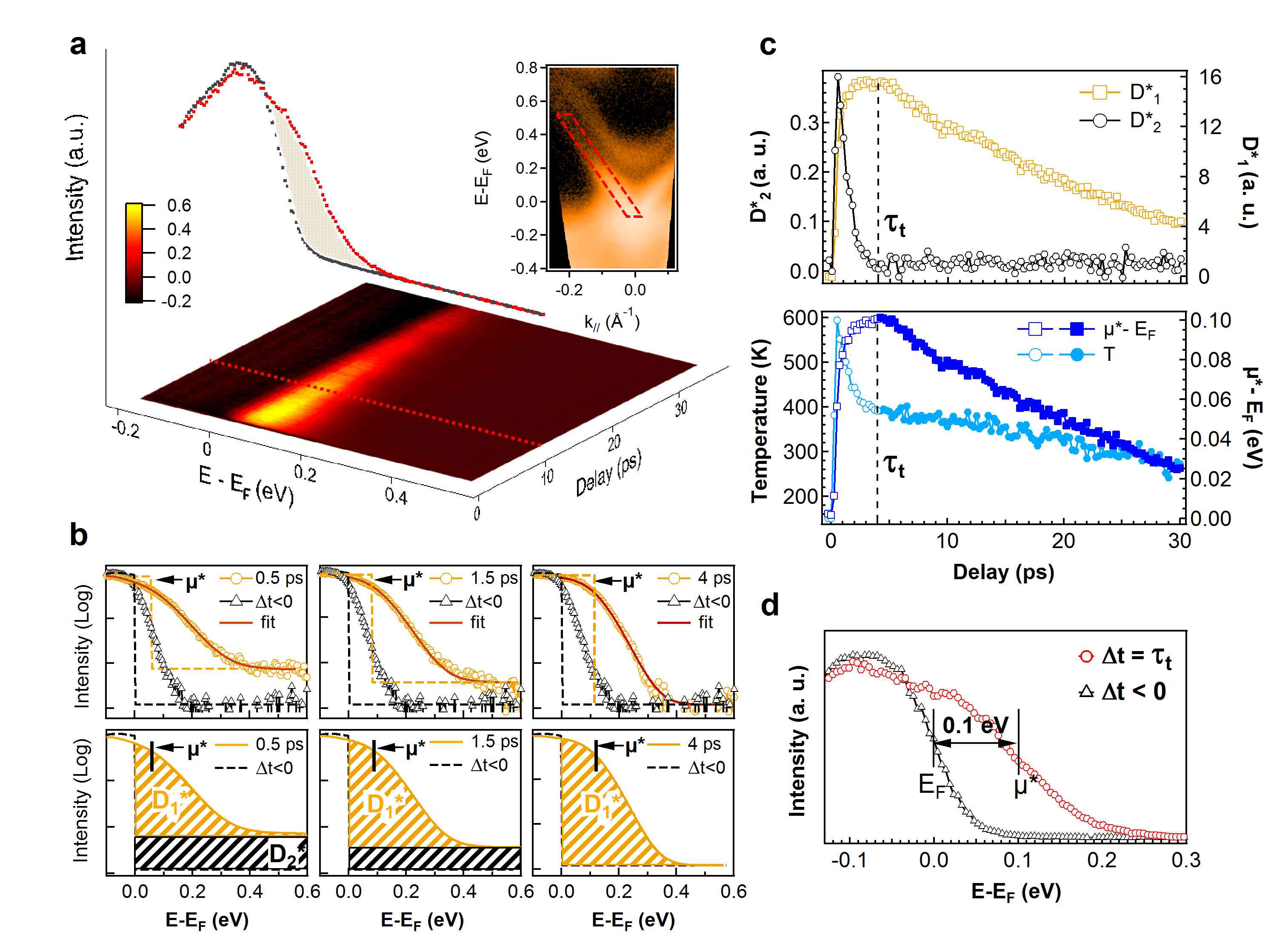}\\
  \caption{
{\bf Transient electronic temperature and chemical potential of non-equilibrium Dirac fermions}. (a) Contour plot of the energy-time distribution of transient hot carriers in
the surface Dirac cone of p-type Bi$_{2.2}$Te$_3$ at 130 K. The two spectra above the contour plot are two representative momentum-selected EDC's extracted from the integration area indicated in the inset, for $\Delta t$ = 10 ps and $\Delta t <$ 0 ps, respectively; the excess electron population is represented by the shaded area between them, and it corresponds to the color scale in the 2D plot. (b) Top panels: momentum-selected EDC's at $\Delta t$ = 0.5, 1.5 and 4 ps with relative fitting curves, and their difference with respect to the EDC taken at $\Delta t <$ 0 ps; bottom panels: visualisation for the various time delays of $D_1$* and $D_2$*, parameters used to fit the transient electronic states with a Fermi-Dirac distribution as explained in the text. (c) Time evolution of the fitting parameters. (d) Comparison of the momentum-selected EDC's for the Dirac cone at equilibrium and for $\Delta t = \tau_t$
}
\label{Fig6} 
\end{figure*}


\begin{thebibliography}{99}

\bibitem{Geim2007} Geim, A.K., and Novoselov, K.S. The rise of graphene, {\it Nature Mater.}, {\bf 6}, 183 (2007).
\bibitem{Hasan2010} Hasan, M.Z., and Kane, C.L. Topological Insulators, {\it Rev. Mod. Phys.} {\bf 82}, 3045 (2010).
\bibitem{Qi2011} Qi, X.-L., and Zhang, S.-C. Topological Insulators and Superconductors, {\it Rev. Mod. Phys.} {\bf 83}, 1057 (2011).
\bibitem{Pesin2012} Pesin D., and MacDonald, A.H. Spintronics and pseudospintronics in graphene and topological insulators, {\it Nature Mater.}, {\bf 11}, 409 (2011). 

\bibitem{Xia2009} Xia, F. et al. Ultrafast graphene photodetector, {\it Nature Nanotech.}, {\bf 4}, 839 (2009).
\bibitem{Sun2012} Sun, D. et al. Ultrafast hot-carrier-dominated photocurrent in graphene, {\it Nature Nanotech.}, {\bf 7}, 114 (2012). 
\bibitem{Winnerl2011} Winnerl, S. et al. Carrier relaxation in epitaxial graphene photoexcited near the Dirac point, {\it Phys. Rev. Lett.}, {\bf 107}, 237401 (2011). 
\bibitem{Li2012} Li, T. et al. Femtosecond population inversion and stimulated emission of dense Dirac fermions in graphene, {\it Phys. Rev. Lett.}, {\bf 108}, 167401 (2012). 
\bibitem{Gilbertson2012} Gilbertson, S. et al. Tracing Ultrafast Separation and Coalescence of Carrier Distributions in Graphene with Time-Resolved Photoemission, {\it J. Phys. Chem. Lett.}, {\bf 3}, 64 (2012). 

\bibitem{Sobota2013} Sobota, J.A. et al. Direct optical coupling to an unoccupied Dirac surface state in the topological insulator Bi$_{2}$Se$_3$, {\it Phys. Rev. Lett.} {\bf 111}, 136802 (2013). 
\bibitem{Sobota2012} Sobota, J.A. et al. Ultrafast optical excitation of a persistent surface state population in the topological insulator Bi$_{2}$Se$_3$, {\it Phys. Rev. Lett.} {\bf 108}, 117403 (2012).
\bibitem{Hajlaoui2012} Hajlaoui, M. et al. Ultrafast surface carrier dynamics in the topological insulator Bi$_{2}$Te$_3$, {\it Nano Lett.} {\bf 12}, 3532 (2012).
\bibitem{Wang2012} Wang, Y.H., et al. Measurement of intrinsic Dirac fermion cooling on the surface of the topological insulator Bi$_{2}$Se$_3$ using time-resolved and angle-resolved photoemission spectroscopy, {\it Phys. Rev. Lett} {\bf 109}, 127401 (2012).
\bibitem{Crepaldi2012} Crepaldi, A., et al. Ultrafast photodoping and effective Fermi-Dirac distribution of the Dirac particles in Bi$_{2}$Se$_3$, {\it Phys. Rev. B} {\bf 86}, 205133 (2012).
\bibitem{Crepaldi2013} Crepaldi, A., et al. Evidence of reduced surface electron-phonon scattering in the conduction band of Bi$_{2}$Se$_3$ by nonequilibrium ARPES, {\it Phys. Rev. B} {\bf 86}, 205133 (2013).

\bibitem{Bokor1989} Bokor, J. Ultrafast dynamics at semiconductor and metal surfaces {\it Science} {\bf 246}, 1130 (1989).
\bibitem{Halas1989} Halas, N.J. and Bokor, J. Surface Recombination on the Si(111)2x1 Surface, {\it Phys. Rev. Lett.} {\bf 62}, 1679 (1989). 
\bibitem{Marsi1997} Marsi, M. et al. Surface states and space charge layer dynamics on  Si(111)2x1: a Free Electron Laser - synchrotron radiation study, {\it Appl. Phys. Lett.} {\bf 70}, 895 (1997). 
\bibitem{Toben2005} Toben, L. et al. Femtosecond transfer dynamics of photogenerated electrons at a surface resonance of reconstructed InP(100), {\it Phys. Rev. Lett.}  {\bf 94}, 067601 (2005). 

\bibitem{Zhang2009} Zhang, H., et al. Topological insulators in Bi$_{2}$Se$_3$, Bi$_{2}$Te$_3$ and Sb$_{2}$Te$_3$ with a single Dirac cone on the surface, {\it Nature Phys.} {\bf 5}, 438 (2009).
\bibitem{Yazyev2010} Yazyev, O. et al. Spin polarization and transport of surface states in the topological insulators Bi$_{2}$Se$_3$ and Bi$_{2}$Te$_3$ from first principles, {\it Phys. Rev. Lett.} {\bf 105}, 266806 (2010).
\bibitem{Hsieh2008} Hsieh, D., et al. A topological Dirac insulator in a quantum spin Hall phase, {\it Nature} {\bf 452}, 970 (2008).
\bibitem{Chen2009} Chen, Y.L., et al. Experimental Realization of a Three-Dimensional Topological Insulator, Bi$_{2}$Te$_3$, {\it Science} {\bf 325}, 178 (2009).

\bibitem{Jia2012} Jia, S., et al. Defects and high bulk resistivities in the Bi-rich tetradymite topological insulator Bi$_{2+x}$Te$_{2-x}$Se, {\it Phys. Rev. B} {\bf 86}, 165119 (2012).
\bibitem{Xiong2012} Xiong, J., et al. High-field Shubnikov-de Haas oscillations in the topological insulator Bi$_{2}$Te$_{2}$Se, {\it Phys. Rev. B} {\bf 86}, 045314 (2012).
\bibitem{Ji2012} Xiong, J., et al. Bi$_{2}$Te$_{1.6}$Se{1.4}: a Topological Insulator in the tetradymite family, {\it Phys. Rev. B} {\bf 85}, 201103 (2012).
\bibitem{Niesner2012} Niesner, D., et al. Unoccupied topological states on bismuth chalcogenides, {\it Phys. Rev. B} {\bf 86}, 205403 (2012).
\bibitem{Mi2013} Mi, J., et al. Phase separation and bulk p-n transition in single crystals of Bi$_{2}$Te$_{2}$Se topological insulator, {\it Adv. Mater.} {\bf 25}, 889 (2013).


\bibitem{Rodolakis2009} Rodolakis, F. et al. Quasiparticles at the Mott Transition in V$_{2}$O$_3$: Wave Vector Dependence and Surface Attenuation, {\it Phys. Rev. Lett.},  {\bf 102}, 066805 (2009). 
\bibitem{Kronik1999} Kronik,L. and Shapira, Y. Surface photovoltage phenomena: theory, experiment and applications, {\it Surf. Sci. Reports} {\bf 37}, 1 (1999).
\bibitem{Marsi1998} Marsi, M. et al. Surface photovoltage in semiconductors under pulsed optical excitation, {\it J. Electron Spectrosc. Relat. Phenom.} {\bf 94}, 149 (1998).
\bibitem{Sze} Sze, S.M. Physics of Semiconductor Devices, John Wiley and Sons. 

\bibitem{Lisowski2004} Lisowski, M. et al. Ultrafast dynamics of electron thermalization, cooling and transport effects in Ru(001), {\it Appl. Phys. A} {\bf 78}, 165 (2004).
\bibitem{Hicks2013} Hicks, J. et al. A wide bandgap metal-semiconductor-metal nanostructure made entirely from graphene, {\it Nature Phys.} {\bf 9}, 49 (2013).
\bibitem{Hsieh2009} Hsieh, D. et al. A tunable topological insulator in the spin helical Dirac transport regime, {\it Nature} {\bf 460}, 1101 (2009).
\bibitem{Bianchi2010} Bianchi, M. et al. Coexistence of the topological state and a two-dimensional electron gas on the surface of Bi$_{2}$Se$_3$, {\it Nat. Commun.},  {\bf 1}, 128 (2010).
\bibitem{Faure2012} Faure, J. et al. Full characterization and optimization of a fs ultraviolet laser source for time and angle-resolved photoemission, {\it Rev. Sci. Instrum.}, {\bf 83}, 043109 (2012).




\end{thebibliography}
\end{document}